\documentclass[aps,pra,reprint,twocolumn,floatfix,notitlepage, floatfix,superscriptaddress,longbibliography]{revtex4-1}

\usepackage[T1]{fontenc}
\usepackage{lmodern}
\usepackage[utf8]{inputenc}
\usepackage[english]{babel}

\usepackage{microtype}
\usepackage{indentfirst}	

\usepackage{amsthm, amssymb, amstext, amsmath, bbm}
\usepackage{dsfont}
\usepackage{cancel}

\usepackage{float}
\usepackage{tikz}
\usepackage{soul}

\usepackage{epstopdf}
\usepackage{blkarray}
\usepackage{booktabs}
\usepackage{braket}

\usepackage{xcolor}

\usepackage{array}

\definecolor{darkred}{rgb}{0.5,0,0}
\definecolor{darkgreen}{rgb}{0,0.5,0}
\definecolor{darkblue}{rgb}{0,0,0.5}

\usepackage[colorlinks]{hyperref}
\hypersetup{colorlinks,
linkcolor=darkred,
filecolor=darkgreen,
urlcolor=darkblue,
citecolor=darkgreen}

\renewcommand{\eqref}[1]{\mbox{Eq.~(\ref{#1})}}

\newcommand{\be}{\begin{equation}}
\newcommand{\ee}{\end{equation}}
\newcommand{\bea}{\begin{eqnarray}}
\newcommand{\eea}{\end{eqnarray}}

\usepackage[colorlinks]{hyperref}
\hypersetup{%
	plainpages=true,
	breaklinks=true,
	hypertexnames=false,
	pageanchor=true,
	colorlinks=true,
	linkcolor={blue},
	citecolor={red},
	urlcolor={blue},
	anchorcolor={black}
}

\usepackage{natbib}
\usepackage{graphicx}

\newcommand{\DD}{\mathcal{D}}
\newcommand{\rhot}{\hat{\rho}(t)}
\newcommand{\de}{{\rm d}}
\usepackage{braket}

\newcommand{\expect}[1]{\left\langle #1 \right\rangle}

\begin{document}

\title{Dissipative state transfer and Maxwell's demon in single quantum trajectories:
Excitation transfer between two noninteracting qubits via
unbalanced dissipation rates}

\author{Fabrizio Minganti}
\email{fabrizio.minganti@riken.jp}
\affiliation{Theoretical Quantum Physics Laboratory, RIKEN, Wako-shi, Saitama 351-0198, Japan}
\author{Vincenzo Macr\`{i}}
\email{vincenzo.macri@riken.jp}
\affiliation{Theoretical Quantum Physics Laboratory, RIKEN, Wako-shi, Saitama 351-0198, Japan}
\author{Alessio Settineri}
\affiliation{Dipartimento di Scienze Matematiche e Informatiche, Scienze Fisiche e  Scienze della Terra, Universit\`{a} di Messina, I-98166 Messina, Italy}
\author{Salvatore Savasta}
\affiliation{Theoretical Quantum Physics Laboratory, RIKEN, Wako-shi, Saitama 351-0198, Japan}
\affiliation{Dipartimento di Scienze Matematiche e Informatiche, Scienze Fisiche e  Scienze della Terra, Universit\`{a} di Messina, I-98166 Messina, Italy}
\author{Franco Nori}
\affiliation{Theoretical Quantum Physics Laboratory, RIKEN, Wako-shi, Saitama 351-0198, Japan}
\affiliation{Physics Department, The University of Michigan, Ann Arbor, Michigan 48109-1040, USA}

\begin{abstract}
We introduce a protocol to transfer excitations between two noninteracting qubits via purely dissipative processes (i.e., in the Lindblad master equation there is no coherent interaction between the qubits). The fundamental ingredients are the presence of collective (i.e. nonlocal) dissipation and unbalanced local dissipation rates (the qubits dissipate at different rates).
The resulting quantum trajectories show that the measurement backaction changes the system wave function and induces a passage of the excitation from one qubit to the other.
While similar phenomena have been witnessed for a non-Markovian environment, here the dissipative quantum state transfer is induced by an update of the observer knowledge of the wave function in the presence of a Markovian (memoryless) environment---this is a single quantum trajectory effect.
Beyond single quantum trajectories and postselection, such an effect can be observed by histogramming the quantum jumps along several realizations at different times. 
By investigating the effect of the temperature in the presence of unbalanced local dissipation, we demonstrate that, if appropriately switched on and off, the collective dissipator can act as a Maxwell's demon.
These effects are a generalized measure equivalent to the standard projective measure description of quantum teleportation and Maxwell's demon.
They can be witnessed in state-of-the-art setups given the extreme experimental control in, e.g., superconducting qubits or Rydberg atoms.
\end{abstract}

\maketitle

\section{Introduction}

An open quantum system is characterized by dissipative processes which, for a weak and Markovian environment, are captured by a Lindblad master equation (LME) \cite{BreuerBookOpen,SettineriPRA2018}. According to measurement theory \cite{Wiseman_BOOK_Quantum,Gardiner_BOOK_Quantum}, the LME admits a fascinating interpretation.
Indeed, it can be recast as the result of a series of continuous and unread ``generalized measures'', called positive operator-valued measures (POVMs)\cite{peres1993quantum}. The action of the environment can be modelled as detectors that continuously measure the output field of the system whose outcomes are unknown to an observer.
However, when an observer reads such a measurement output, rather than a description by LME, the system is characterized by a stochastic evolution, called \textit{quantum trajectory} \cite{CarmichaelPRL93,DalibardPRL92,Molmer1993,DaleyAdvancesinPhysics2014}.

Experimentally, there exist different ways in which the output fields can be monitored.
This translates into different evolution functions for the quantum trajectories \cite{Plenio1998,Haroche_BOOK_Quantum}, e.g.: (i) Non-Hermitian continuous time evolution interrupted by abrupt changes in the wave function due to quantum jumps. This is the widely used counting trajectory (since one counts the number of quantum jumps), also known as Monte Carlo wave function method \cite{Molmer1993};
(ii) Continuous stochastic infinitesimal changes of the wave function due to a noise term. This is a quantum state diffusion method, often called homodyne (heterodyne) trajectories in the photonic context \cite{Gisin1992,gisin1993,GisinAA1993,Percival2002}.
However, no matter the details at the single trajectory level, the average behavior over infinitely many quantum trajectories always recovers the result of the LME.

In the following, we will mainly focus on counting trajectories. For photons escaping an electromagnetic resonator, such a method describes a set of ideal photodetectors, performing POVMs for all the output fields of the system, resulting in a series of annihilation operators which act at random times \cite{Wiseman_BOOK_Quantum,Gardiner_BOOK_Quantum}. 
Within this description, the continuous monitoring of the detectors affects the wave function even when no excitation is lost to the environment \cite{Haroche_BOOK_Quantum}.
Indeed, the very fact of knowing that no \textit{quantum jump} took place modifies the wave function.

Normally, dissipation is regarded as a source of classicality \cite{ZurekRMP03}. Indeed, a quantum system in contact with its \textit{classical} environment rapidly loses its quantum features, such as entanglement.
Several works demonstrated that two qubits strongly interacting with a non-Markovian common bath display revival of entanglement \cite{PlenioPRA99,LiPRB08}, possibly enhanced by the presence of measurement protocols (quantum Zeno effect) \cite{ManiscalcoPRL2008,LiJP09,MaPRA2012}, and revealing nontrivial dynamics \cite{FrancicaPRA09,MazzolaPRA2009}.

In all these cases, the presence of a non-Markovian environment play a fundamental role: the excitations lost in the environment can induce effects on the system itself.
In this regard, Markovian dissipation would suppress quantum effects since all information is immediately lost into the environment.
However, through correctly engineering the environment and by considering quantum trajectories, it is possible to realize such quantum processes \cite{ReiterPhysRevLett2016}.
A pivotal example is the entanglement of two superconducting qubits, separated by more than a meter of coaxial cable, by designing a joint measurement of the two qubits \cite{RochPRL2014}.

Although a single trajectory is a Markovian stochastic process, it can significantly deviate from the LME result, indicating the importance of the dynamics on its past.
Indeed, the initial condition does not determine the system state at a certain time, but one needs to keep track of all the jump events to correctly describe the system.
In this regard, the ``observer'' of a quantum trajectory plays the role of a classical memory by keeping track of the jumps.
As such, within a single trajectory the continuous monitoring can hinder the emergence of classical states \cite{GalveSciRep16}.

In this work, we consider two undriven and noninteracting qubits whose quantum trajectories stems from a LME (i.e., for a bath within a Born and Markov approximation \cite{BreuerBookOpen,LidarLectureNotes}) in the presence of local and collective dissipation.
As described in Ref.~\cite{ShammahPRA18}, dissipation can be (engineered to be) both ``local'' and ``collective'', meaning that the LME cannot be brought to a form where all the quantum jump operators involve only single qubit jumps. For instance, this is the case of the collective dissipation of the Dicke model, where only the uniform modes are dissipated and the physics can be reduced to one of the Dicke ladders \cite{DickePhysRev1954,ShammahPRA18}.

By tuning the dissipation rates, we demonstrate that it is possible not only to generate entanglement in the system, but also to \textit{transfer a quantum excitation between the two noninteracting qubits via purely dissipative processes}. 
We stress that the Lindblad master equation describing the two qubits once the environment has been traced out does not contain any Hamiltonian interaction between the qubits (contrary to the effective coupling in, e.g., giant atoms \cite{KockumPRL18, KannanNat20}), but it is only the effect of dissipation that introduce such a coupling.
In the same way in which a quantum teleportation protocol relies on the nonlocality of quantum mechanics, here the key to such a state transfer is the presence of engineered \textit{nonlocal dissipative processes} in the form of collective dissipation.
While in a teleportation protocol a ``standard'' projection-valued measures (PVM) collapses the entangled wave function instantaneously, here the POVM induces the transfer of an excitation from one qubit to another.
If no quantum jump happens, the information collected along the trajectory via the POVM progressively and continuously projects the wave function (i.e., it is a weak measurement \cite{BrunAJP02}). 
We call such a process a \textit{dissipative quantum state transfer}, since both the dissipative and quantum nature of the system are necessary ingredients to ensure the transfer.  
\textit{This event is purely quantum and dissipative in nature, and cannot be explained by a purely Hamiltonian or classical theory}.

Such a protocol can also be realized by correctly postselecting those trajectories where no quantum jump happened \cite{NaghilooNatPhys19,MingantiPRA2020}. Furthermore, the statistics of quantum jumps in the two qubits (even without continuously monitoring the collective jumps of the system) bear witness to the quantum state transfer process.

We show that, surprisingly, this process can also happen in the presence of thermal fluctuations. By controlling the collective dissipation it is possible to obtain a flow of energy from a bath to one at a higher temperature. This is a transfer of ``heat'' at the expense of information and not of work. As such, this protocol is that of a Maxwell's demon \cite{MaruyamaRMP09,leff2014book}, where the collective bath plays the role of a door which can be appropriately opened or closed to ensure the cooling of the cold bath \cite{NajeraPRR20}.

The structure of the article is the following.
In Sec.~\ref{Sec:Transfer_protocol} we introduce the system and the transfer protocol in the zero temperature case.
In particular, in Sec.~\ref{Sec:discussion_jumpless_evolution} we discuss the effective Hamiltonian of the system describing a jumpless evolution upon a counting protocol.
In Sec.~\ref{Sec:Histograms_jumps} we show that the quantum state transfer is witnessed also by the statistics of only the local quantum jumps (i.e., without having to explicitly measure the collective emissions).
In Sec.~\ref{Sec:Post_selection} we prove that it is possible, for this specific zero-temperature case, to obtain the results of the quantum trajectory by post-selection.
In Sec.~\ref{Sec: Maxwell_Daemon} we propose an all-dissipative Maxwell's demon by properly switching on the collective dissipation.
We draw our conclusions in Sec.~\ref{Sec:Conclusion}.

\section{Dissipative state transfer protocol}
\label{Sec:Transfer_protocol}

We consider two noninteracting dissipative qubits.
Their Hamiltonian reads ($\hbar=1$):
\begin{equation}
\label{Eq:Hamiltonian}
    \hat{H} = \frac{\omega_1}{2} \hat{\sigma }_z^{(1)}+ \frac{\omega_2}{2} \hat{\sigma}_z^{(2)}.
\end{equation}
where $\hat{\sigma}_z^{(j)}$ is the Pauli $z$ operator of the $j$-th qubit.
We indicate a state as, e.g., $\ket{e,g}$, meaning that qubit 1 is in the excited state and qubit 2 is in the ground state. 

Each qubit interacts with the environment, resulting in a dissipation of their excitations. For a weak and Markovian environment, the evolution of the system density matrix $\rhot$ is captured by a LME, reading 
\begin{equation}\label{Eq:lindblad}
    \partial_t \rhot =  - i [\hat{H}, \rhot] +  \sum_{\mu} \gamma_\mu \DD[\hat{J}_\mu] \rhot,
\end{equation}
where the dissipator $\DD[\hat{J}_\mu]$ of the jump operator $\hat{J}_\mu$ acts via
\begin{equation}
    \DD[\hat{J}_\mu] \rhot = \hat{J}_\mu \rhot \hat{J}_\mu^\dagger -\frac{\hat{J}_\mu^\dagger \hat{J}_\mu \rhot + \rhot \hat{J}_\mu^\dagger \hat{J}_\mu}{2},
\end{equation}
at a rate $\gamma_\mu$.

In our case, we assume that each qubit dissipates its excitation into the environment both {\emph {locally}} and {\emph {collectively}}, using the terminology of Ref.~\cite{ShammahPRA18}.
Thus, the dissipators read:
\begin{subequations}
\begin{equation}
 \gamma_1 \DD\left[\hat{\sigma}_{-}^{(1)} \right]   ,
\end{equation}
\begin{equation}
    \gamma_2 \DD\left[\hat{\sigma}_{-}^{(2)} \right] ,
\end{equation}
\begin{equation}
     \gamma_c \DD\left[\frac{\hat{\sigma}_{-}^{(1)}+ \hat{\sigma}_{-}^{(2)}}{\sqrt{2}}\right],
\end{equation}
\end{subequations}
where $\hat{\sigma}_\pm^{(j)}=(\hat{\sigma}_x^{(j)} \pm i \hat{\sigma}_y^{(j)})/2$ is the lowering (raising) operator of the $j$-th qubit.
While the first two dissipators characterize the dynamics of noninteracting qubits, the collective dissipation and its rate $\gamma_c$ naturally emerge when the qubits are sufficiently close to each other with respect to the typical wavelength of the electromagnetic reservoir (see the discussion in, e.g., \cite{MacriPRA20}). 
Furthermore, the collective dissipation can be engineered for noninteracting nonlinear cavities, as discussed in Ref.~\cite{MamaevQuantum18}, such as circuit-QED systems \cite{Gu2017}.
For strong (i.e., infinite) photon-photon interaction, the physics of the nonlinear resonators becomes that of a two-level system  \cite{Carusotto_RMP_2013_quantum_fluids_light,CarusottoPRB2005,BirnbaumNat05,DelteilNatMat19,LangPRL11}.

One can assume that all emitted excitations are perfectly monitored by detectors (or, more generally, measurement instruments), one for each jump operator.
As such, the state of the system is perfectly known and one speaks of \emph{quantum trajectories} \cite{Molmer1993,CarmichaelPRL93,Wiseman_BOOK_Quantum,DaleyAdvancesinPhysics2014}.
The LME is transformed into a stochastic Schr\"odinger equation known as \emph{counting} quantum trajectory \cite{Wiseman_BOOK_Quantum}. 
The equation of motion becomes
\begin{equation}
\label{Eq:SSE}
\begin{split}
\de \ket{\psi(t)}&=  \left[\sum_{\mu} \gamma_\mu \de N_{\mu}(t)\left(\frac{\hat{J}_{\mu}}{\sqrt{\braket{\hat{J}^\dagger_{\mu} \hat{J}_{\mu} }}} - \mathds{1}\right) \right.  \\
& \qquad - i  \de t \hat{H}_{\rm eff}  \Bigg]\ket{\psi(t)},
\end{split}
\end{equation}
where $\de N_{\mu}(t)$ is a stochastic variable whose value is either 0 or 1
with probability \begin{equation}
p\left[\de N_{\mu}(t)=1\right] = \gamma_\mu \de t \expect{\psi(t)\left| \hat{J}_\mu \right|\psi(t)}.
\end{equation} 
$N_{\mu}(t)$ counts the occurrences of a jump induced by the operator $\hat{J}_\mu$.
Moreover, $\hat{H}_{\rm eff}$ is the effective Hamiltonian and reads
\begin{equation}\label{Eq:Effective_Hamiltonian_counting}
\hat{H}_{\rm eff} =\hat{H}  - i \sum_\mu \gamma_{\mu}\left(  \frac{\hat{J}_{\mu}^\dagger \hat{J}_{\mu}}{2} -    \frac{\braket{\hat{J}_{\mu}^\dagger \hat{J}_{\mu}}}{2}\right).
\end{equation}

Within this description, the LME can be seen as the average over an infinite number of quantum trajectories. Therefore, each behaviour of the LME can be witnessed by any quantum trajectory approach \cite{Wiseman_BOOK_Quantum}. However, some of the behaviours at the single quantum trajectory level are washed out by the averaging process \cite{BartoloEPJST17,RotaNJP18,MunozPRA19}.
Indeed, single trajectories ``contain more information'' since one eventually knows  that a quantum jumps happened and how excitations are lost to the environment. In this regard, single quantum trajectories mimic the result of single runs of an idealized experiment.

\subsection{Evolution without jumps}
\label{Sec:discussion_jumpless_evolution}

To obtain the desired effect of state transfer from qubit 1 to qubit 2, having initialized the system in qubit 1 ($\ket{\psi(0)}=\ket{e,g}$), we have to ensure the following conditions: (i) No quantum jump takes place (otherwise, the excitation is lost into the environment); (ii) The collective dissipation is different from zero, $\gamma_c\neq 0$. Moreover, we can obtain an excitation swap if (iii) the dissipation of qubit 1 is larger than that of qubit 2,  $\gamma_1> \gamma_2$.

In the absence of quantum jumps, the evolution of the system is dictated solely by the effective Hamiltonian in \eqref{Eq:Effective_Hamiltonian_counting}.
Thus, the effect of \emph{not} measuring modifies the system wave function via the renormalization induced by the non-Hermitian terms of $\hat{H}_{\rm eff}$ in \eqref{Eq:Effective_Hamiltonian_counting} \cite{Molmer1993}.
Generally speaking, given a superposition of ``bright'' and ``dark'' states of the dissipation, each time there is no quantum jump the wave function becomes more populated by the dark state.

In the main text, we will suppose that the two qubits have the same energy, i.e., $\omega_1=\omega_2$.
The role of the qubit frequencies in the dynamics is discussed in Appendix~\ref{Appendix_A}.

\subsubsection{Environment backation for a single qubit}

In the case of a single qubit, the environment back-action through the continuous measure is clear, as described in Refs.~\cite{Molmer1993}.
The Hamiltonian is
\begin{equation}
\hat{H}=\frac{\omega}{2} \hat{\sigma}_z    
\end{equation}
and the system is initialized in a superposition state $(\ket{e}+\ket{g})/{\sqrt{2}}$.
The system is subject to a dissipation
\begin{equation}
    \gamma \mathcal{D}[\hat{\sigma}_{-}]=\gamma \mathcal{D}[\ket{g}\bra{e}].
\end{equation} 
In this case, the state $\ket{e}$ is the bright state and $\ket{g}$ is the dark one. 
Since the effective Hamiltonian now reads
$\hat{H}_{\rm eff}= \omega/2 \hat{\sigma}_z - i \gamma/2 \, \hat{\sigma}_+ \hat{\sigma}_-$,\textit{ each time there is no quantum jump} [$\de N(t) =0$] \textit{the weight of the state} $\ket{g}$ \textit{increases}.
Equivalently, the observer gains information on the wave function, and the fact that there was no jump makes the state $\ket{g}$ more probable.
Thus, the non-Hermitian part of \eqref{Eq:Effective_Hamiltonian_counting} continuously affects the wave function, which approaches the ground state as time passes and no detection happens. This can be seen as a Bayesian update of the wave function \cite{ChantasriPRX16}.

\subsubsection{Environment backation for two qubits with collective dissipation}

\begin{figure}
    \centering
    \includegraphics[width=0.94 \linewidth]{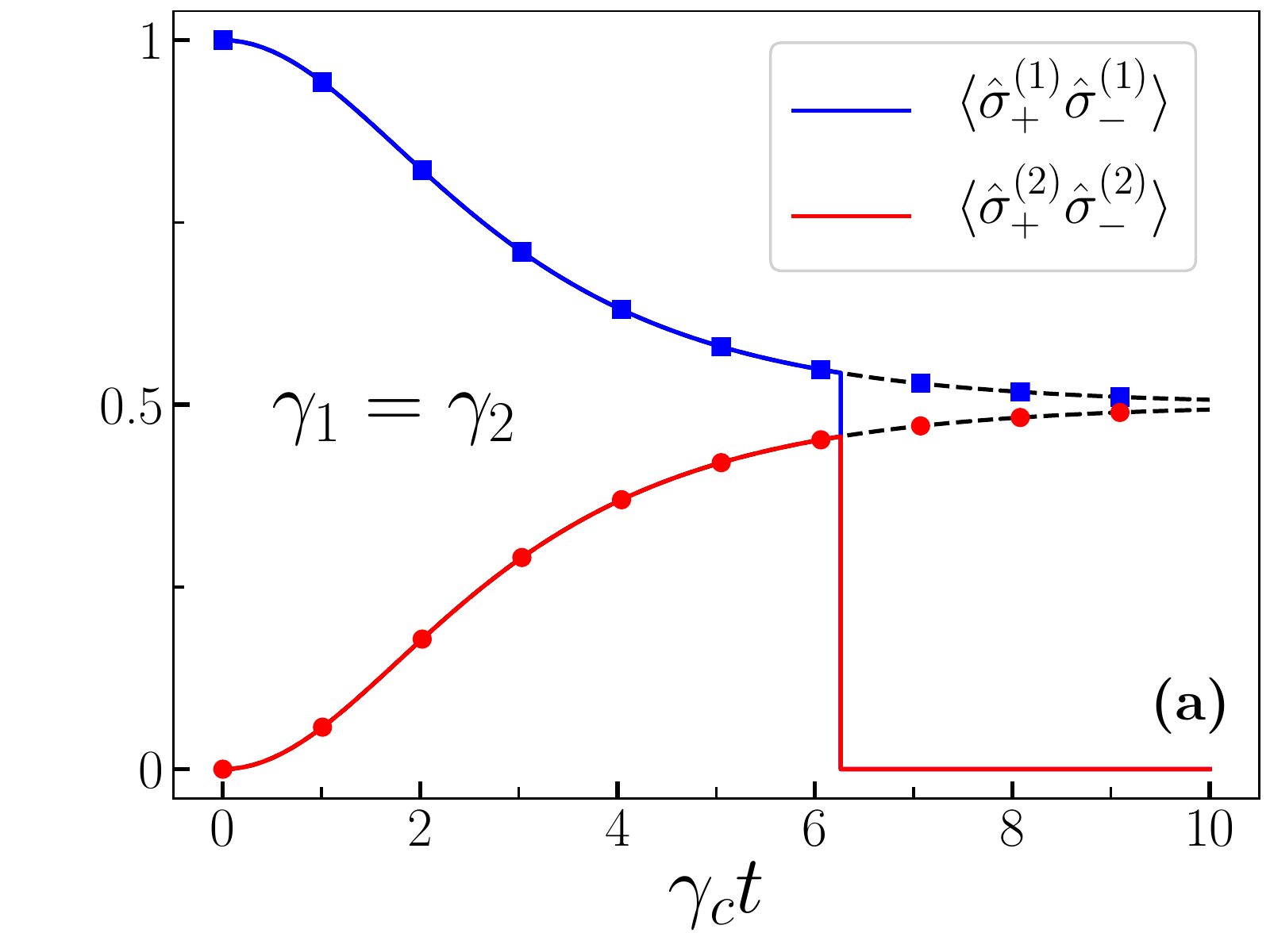} \\
    \includegraphics[width=0.94 \linewidth]{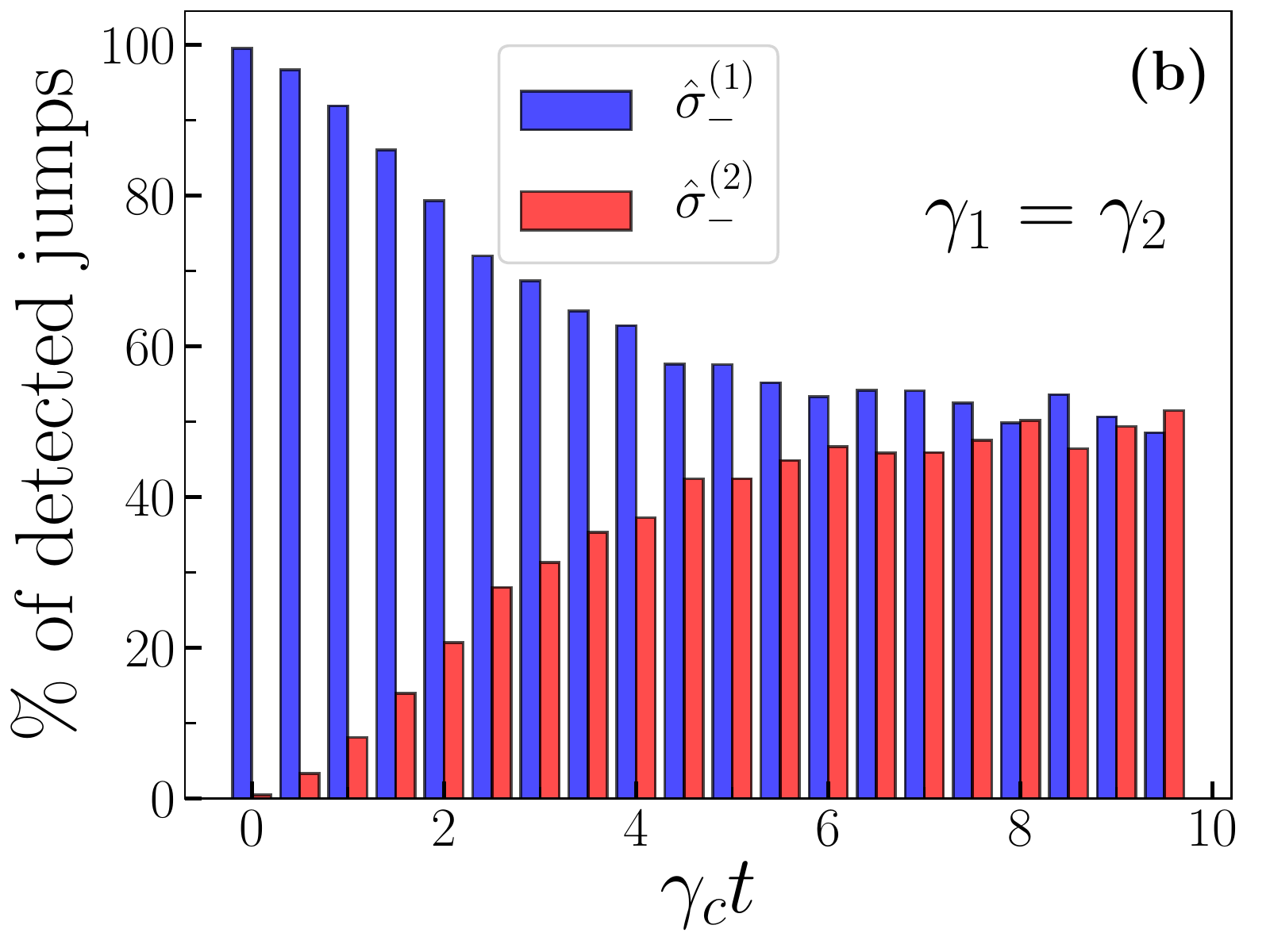}
    \caption{ (a) As a function of time, $\hat{\sigma}_+^{(1)}\hat{\sigma}_-^{(1)}$ (qubit 1, blue curve) and $\hat{\sigma}_+^{(2)}\hat{\sigma}_-^{(2)}$ (qubit 2, red curve) for an ideal counting quantum trajectory in the presence of collective dissipation and local dissipation identical for qubits 1 and 2. As time passes, the wave function tends towards the Bell state $\ket{\Psi^-}$, until a quantum jump happens at time $\gamma_c t \simeq 6$.
    The black dashed lines represent the analytical results of \eqref{Eq:Transfer_ideal}, while the blue (red) circles (squares) indicate the results of the postselection in \eqref{Eq:postselection} for qubit 1 (2).
    (b) As a function of time and for a single trajectory, the percentage of local jumps $\hat{\sigma}_-^{(1)}$ happening for qubit 1 (blue bars) and $\hat{\sigma}_-^{(2)}$ for qubit 2 (red bars) without keeping track of the collective quantum jumps due to $(\hat{\sigma}_-^{(1)}+\hat{\sigma}_-^{(2)})/\sqrt{2}$. We simulated $10^6$ trajectories.
    Parameters: $\omega_1=\omega_2=10 \gamma_c$ (the relevant condition is $\omega_1=\omega_2$),  $\gamma_1/\gamma_c=0.2,$ $\gamma_2/\gamma_c=0.2$.  The system is always initialized in the state $\ket{e, g}$. }
    \label{fig:no_local_dissipation}
\end{figure}
 \begin{figure}
    \centering
    \includegraphics[width=0.94 \linewidth]{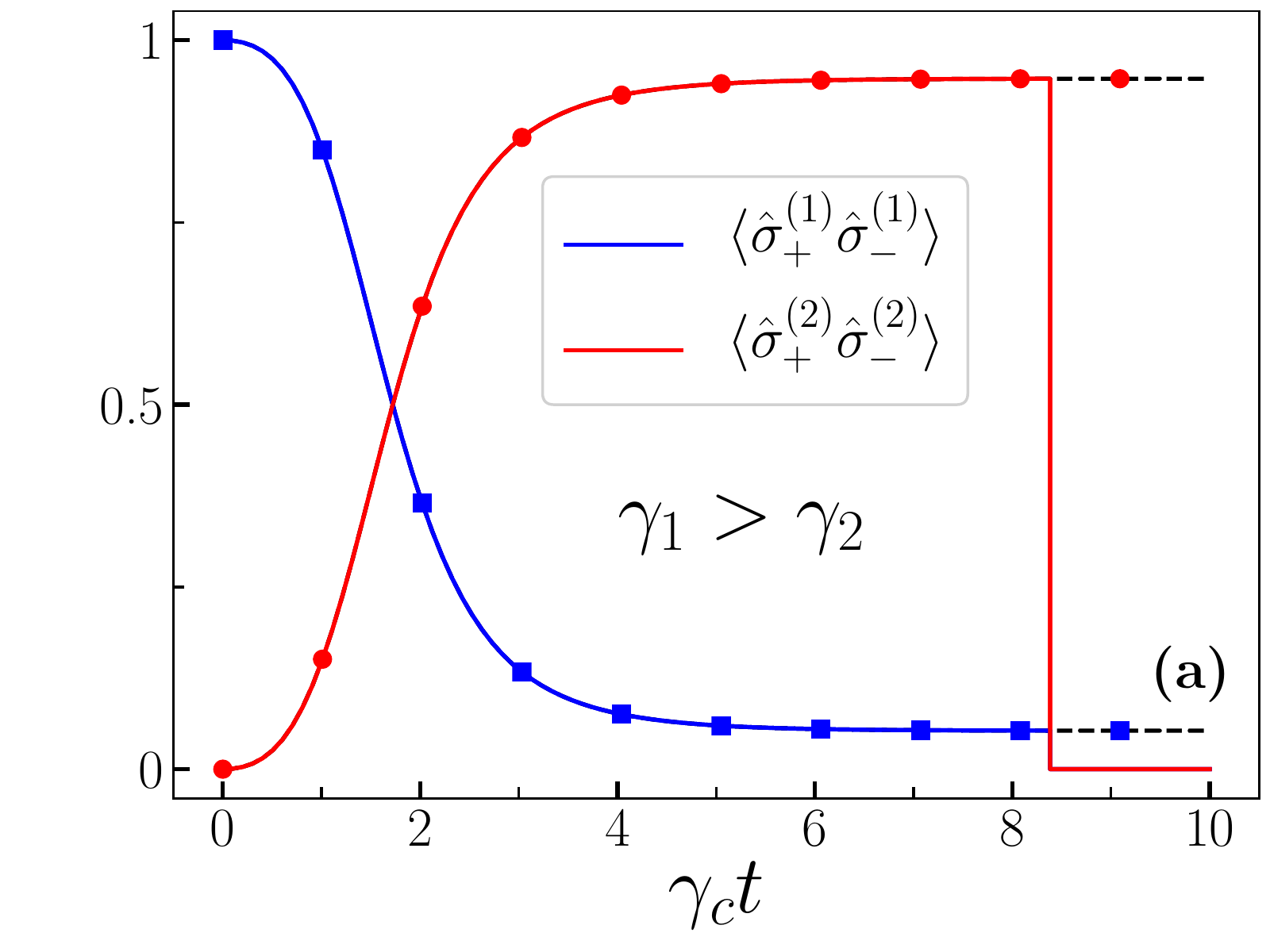}
    \includegraphics[width=0.94 \linewidth]{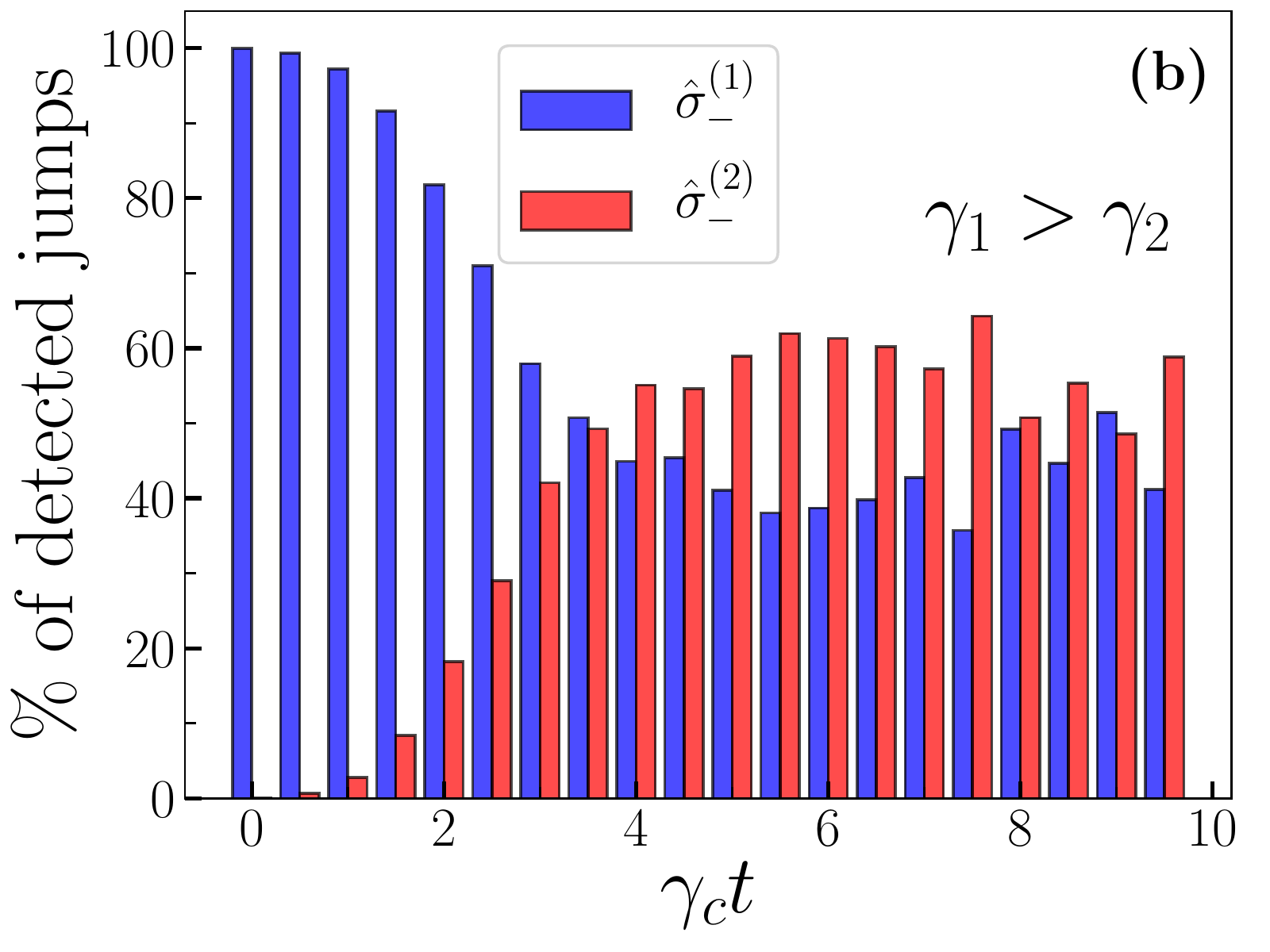}
    \caption{
    (a) As a function of time, the expectation value of $\hat{\sigma}_+^{(1)}\hat{\sigma}_-^{(1)}$ (qubit 1, blue curve) and $\hat{\sigma}_+^{(2)}\hat{\sigma}_-^{(2)}$ (qubit 2, red curve) for an ideal counting quantum trajectory in the presence of collective dissipation and unbalanced local dissipation. While collective disipation creates a Bell state $\ket{\Psi^-}$, the unbalance in local dissipation transforms $\ket{\Psi^-}$ into $\ket{g,e}$. These two processes ensure the excitation swap, until a quantum jump happens at $\gamma_c t \simeq 8$. The black dashed lines represent the analytical results of \eqref{Eq:Transfer_ideal}, while the blue (red) circles (squares) indicate the results of the postselection in \eqref{Eq:postselection} for qubit 1(2).
    (b) As a function of time, the percentage of local jumps $\hat{\sigma}_-^{(1)}$ happening for qubit 1 (red bars) and $\hat{\sigma}_-^{(2)}$ for qubit 2 (blue bars) without keeping track of the collective quantum jumps due to $(\hat{\sigma}_-^{(1)}+\hat{\sigma}_-^{(2)})/\sqrt{2}$. Despite the remarkable difference in the emission rate, the statistics of jumps at long time clearly indicates the quantum excitation swap. The last time intervals are noisy due to the lack of statistics. We simulated $10^6$ trajectories.
    Parameters: $\omega_1=\omega_2=10 \gamma_c$ (the relevant condition is $\omega_1=\omega_2$), $\gamma_1/\gamma_c=2.2,$ $\gamma_2/\gamma_c=0.2$. The system is always initialized in the state $\ket{e, g}$.
    }
    \label{fig:perfect_transfer}
\end{figure}

The collective operator acts in a very similar way on the state $\ket{e,g}$.
In this case, however, there exist multiple bright and dark states, according to the number of excitations in the system.
Indeed, by introducing the Bell states $\ket{\Psi^\pm}=(\ket{e, g} \pm \ket{g, e})/\sqrt{2}$, we have 
\begin{equation}
    \ket{e,g}=\left( \ket{\Psi^+} + \ket{\Psi^-} \right)/\sqrt{2}.
\end{equation}
Since we always initialize the system in $\ket{e,g}$, the collective dissipator reads
\begin{equation}\label{Eq:collective_dissipator_Bell}
   \gamma_c \mathcal{D}\left[\frac{\hat{\sigma}_{-}^{(1)}+ \hat{\sigma}_{-}^{(2)}}{\sqrt{2}}\right] = \gamma_c \mathcal{D}\left[\ket{g,g}\bra{\Psi^+}\right].
\end{equation}
Thus, both $\ket{g,g}$ and $\ket{\Psi^{-}}$ are dark states of the dissipator in \eqref{Eq:collective_dissipator_Bell}, while $\ket{\Psi^{+}}$ is the bright state.

The dynamics of the initial state $\ket{e,g}$ is, therefore, characterized by two possible processes.
If a quantum jump happens, the wave function at time $t$ becomes immediately $\ket{g,g}$ and there is no more dynamics.
Instead, as time passes and when no quantum jump happens, the state $\ket{\Psi^-}$ becomes more probable.
This process is captured by projecting the effective Hamiltonian in \eqref{Eq:Effective_Hamiltonian_counting} onto the one-excitation manifold, which in the presence of only the collective dissipation (up to a constant) reads 
\begin{equation}
   \expect{\Psi^{\pm}|\hat{H}_{\rm eff}|\Psi^{\pm}} = - i \frac{\gamma_c}{2} \ket{\Psi^{+}}\bra{\Psi^{+}}.
\end{equation}
Thus, when no quantum jump happens the coefficient of the state $\ket{\Psi^{+}}$ exponentially decays to zero, and therefore $\ket{\Psi(t)}$ approaches $\ket{\Psi^-}$.
We show this effect in Fig.~\ref{fig:no_local_dissipation}(a), where $\ket{e, g}$ converges towards $\ket{\Psi^{-}}$ before a quantum jump takes place.
The red and blue curves represent the expectation value of $\hat{\sigma}^{(1,\,2)}_+\hat{\sigma}_-^{(1,\,2)}$ obtained from a quantum trajectory simulation.
For $\gamma_c t \simeq 6$ a quantum jump happens and the system loses its excitation and ends up in $\ket{g,g}$.

In other words, condition (i) and (ii) ensure that the system is capable of \textit{producing entanglement along the dissipative dynamics when no quantum jump happens}.

\subsubsection{Environment-induced excitation swap between the qubits}

Condition (iii) states that qubit 1 dissipates faster than qubit 2 ($\gamma_1>\gamma_2$). Suppose now that we have no collective dissipation and we initialize the system in $\ket{\Psi^{\pm}}$.
If no local quantum jump is detected, it becomes more probable to find an excitation qubit 2 because it is less dissipative. Thus, a non-localized state collapses towards the state $\ket{g,e}$.

If the collective and local dissipations act simultaneously according to the conditions (i)-(iii),  there is a transfer of the excitation $\ket{e,g}$ into a state similar to $\ket{g,e}$ via a local and non-local quantum measurement back-action. 
In a time step, the absence of signal from the collective dissipation projects the state towards $\ket{\Psi_-}$ (i.e., the dark state of \eqref{Eq:collective_dissipator_Bell}). Similarly, the absence of signal from any of the individual dissipation processes gradually makes it more likely that the excitation is mostly in the the less dissipative qubit (in our case, qubit 2). When $\gamma_1 = \gamma_2$, instead, the superposition state $\ket{\Psi_-}$ does not change.

We plot this ideal counting trajectory leading to the excitation swap in Fig.~\ref{fig:perfect_transfer}(a).
The red and blue curves represent the expectation value of $\hat{\sigma}^{(1,\,2)}_+\hat{\sigma}_-^{(1,\,2)}$ obtained from a quantum trajectory simulation.
As we see, there is a transfer of the excitation from qubit 1 to qubit 2. For this realization, at time $\gamma_c t \simeq 8 $, a quantum jump happens, thus ending the transfer process. 

\subsubsection{Analytical description of the time evolution}

The exact diagonalization of the effective Hamiltonian in \eqref{Eq:Effective_Hamiltonian_counting} (done in Appendix \ref{Appendix_A}) analytically describe both Fig.~\ref{fig:no_local_dissipation}(a) and Fig.~\ref{fig:perfect_transfer}(a) (the black dashed curves in both figures). The initial state $\ket{e,g}$ will evolve into an unnormalized wave function
\begin{equation}\label{Eq:Transfer_ideal}
\begin{split}
    \ket{\Psi(t)} & =\left[ \eta  \cosh{\left(\frac{\eta t}{4}\right)} - \Delta \gamma\sinh{\left(\frac{\eta t}{4}\right)} \right] \ket{e,g} \\ &
     \qquad - \gamma_c \sinh{\left(\frac{\eta t}{4}\right)}  \ket{g,e},
    \end{split}
\end{equation}
where $\eta= \sqrt{\gamma_c^2 + \Delta \gamma^2}$ and $\Delta \gamma= \gamma_1-\gamma_2$ (we recall that we are considering $\omega_1=\omega_2$).
Furthermore, given a trajectory, the relevant parameter to determine the probability that an excitation is lost is the effective dissipation rate $\Gamma=\gamma_1+\gamma_2+\gamma_c$ (see the Appendix~\ref{Appendix_A}).

From \eqref{Eq:Transfer_ideal}, a complete transfer happens only for $\gamma_c=0$ and $t \to \infty$.
In any other case, the state will be a superposition of $\ket{e,g}$ and $\ket{g,e}$.
Clearly, one has to play with $\Gamma$ to make the protocol less or more efficient and with the ratios of the parameters to witness a transfer.
For example, for the parameters considered here, we have a fidelity $|\braket{g,e|\Psi(t\to \infty)}|^2\simeq 0.95 $, but only 1 trajectory out of 1000 will reach a time $\gamma_c t \simeq 4$.
The choice $\gamma_1=\gamma_c$ and $\gamma_2=\gamma_c/10$ will lead to $|\braket{g,e|\Psi(t\to \infty)}|^2\simeq 0.83$ (i.e., a lesser fidelity), but 15 trajectories out of 1000 will reach a time $\gamma_c t \simeq 4$ (more trajectories can be observed).
Such a high fidelity is impossible without monitoring quantum trajectories or by using different protocols.
Indeed, as demonstrated in Appendix~\ref{Sec:Homodyne}, \textit{if not using single counting trajectories one misses the fundamental quantum backaction allowing the state transfer}.

Finally, we remark that this state transfer is a rare event, but not an impossible one. The interesting point is not the efficiency of the process, but rather that \textit{this event is exquisitely quantum and dissipative in nature, and cannot be explained by a purely Hamiltonian or a classical theory}.

\subsection{Histograms of the local jumps}
\label{Sec:Histograms_jumps}

Witnessing the quantum state transfer requires to collect all the quantum jumps to observe those trajectories where no quantum jump took place.
Although possible, e.g., for superconducting circuits \cite{NaghilooNatPhys19}, this procedure can be difficult, especially when dealing with collective jumps.
Furthermore, as we just discussed, this transfer is a rare event.

The proof of a quantum state transfer can be obtained also  by histogramming the distribution of the \emph{local} quantum jumps as a function of time.
By partitioning the time of observation, and collecting all the jumps happening in that time frame either from $\mathcal{D}[\hat{\sigma}_-^{(1)}]$ or from $\mathcal{D}[\hat{\sigma}_-^{(2)}]$, one can witness the state transfer.
This procedure has the clear experimental advantage of requiring only to detect the local dissipation processes, and the rarity of the event simply requires repeat more time the experiment to obtain a sufficient long-time statistics.

We plot the histogram of the jumps in Fig.~\ref{fig:no_local_dissipation}(b) for the case $\gamma_1=\gamma_2$ (i.e., generation of the Bell state $\ket{\Psi^{-}}$) and in Fig.~\ref{fig:perfect_transfer}(b) for the case $\gamma_1>\gamma_2$ leading to an excitation swap. 
While initially all the detected jumps occur via $\hat{\sigma}_{-}^{(1)}$, as the time passes more and more jumps involve qubit 2. 
In Fig.~\ref{fig:no_local_dissipation}(b) eventually the number of jumps become identical (up to numerical precision), signalling that $\ket{\Psi(t)}\simeq \ket{\Psi^{-}}$. 
On the contrary, in Fig.~\ref{fig:perfect_transfer}(b) the number of jumps of qubit 2 overcomes those of qubit 1.
Notice that, for the considered values, the dissipation in qubit 1 is 11 times higher than that of qubit 2. Consequently, to compare the histogram in Fig.~\ref{fig:perfect_transfer}(b) with the curves in Fig.~\ref{fig:perfect_transfer}(a) one needs to appropriately rescale the jumps statistics.

Finally, we remark an important characteristic of this histogram protocol with respect to finite-efficiency detectors.
Indeed, if the detector for $\mathcal{D}[\hat{\sigma}_-^{(1)}]$ has the same efficiency of that for $\mathcal{D}[\hat{\sigma}_-^{(2)}]$, the jump statistics is unaffected, since both are missing, on average, the same amount of quantum jumps.

\subsection{Postselection of trajectories with no jumps}
\label{Sec:Post_selection}

We observed that the quantum state transfer happens on those counting trajectories where no quantum jump took place for a certain amount of time. 
This amounts to postselect those trajectories where there are no quantum jumps and discarding all the others.  
As discussed in Ref.~\cite{MingantiPRA2020}, post-selection transforms the master equation in a non-Lindblad form described by a hybrid-Liouvillian superoperator which strongly depends on the form of the jump operators, on their unraveling, and on the efficiency of the detectors.
Thus, by considering a perfect photodetector, we can immediately argue that the results of \eqref{Eq:Transfer_ideal} are obtained by postselection.

Therefore, we can show the presence of a quantum state transfer also at the LME level.
Indeed, we can imagine that we simultaneously measure via a PVM the number of excitations in qubit $1$ and 2 at a time $t$, and we consider only those realizations where an excitation has been found in the system \cite{RochPRL2014}.
In this case, the time of the measurement plays a fundamental role since a mixed state into the pure state $\ket{e,g}$, $\ket{g,e}$, or $\ket{g,g}$. 

Within this assumption, the evolution equation of the post-selected density matrix $\hat{\rho}_{\rm PS}(t)$ measured at time $t$ becomes
\begin{equation}\label{Eq:Post_sel_rho}
    \hat{\rho}_{\rm PS}(t) = \frac{\rhot - \ket{g,g}\bra{g,g} \expect{g,g|\rhot|g,g} }{\operatorname{Tr}\left[\left(\hat{\sigma}_+^{(1)}\hat{\sigma}_-^{(1)}+\hat{\sigma}_+^{(2)}\hat{\sigma}_-^{(2)} \right) \rhot \right] }.
\end{equation}
The evolution assumes this simple form because there are no coherences in the density matrix between states with different number of excitations, i.e., $\expect{e,g|\rhot|g,g}=0$.
Moreover, the denominator ensures that, at the measurement time $t_{\rm m}$, 
\begin{equation}
    \operatorname{Tr}\left[\left(\hat{\sigma}_+^{(1)}\hat{\sigma}_-^{(1)}+\hat{\sigma}_+^{(2)}\hat{\sigma}_-^{(2)} \right) \hat{\rho}_{\rm PS}(t_{\rm m}) \right]=1.
\end{equation}
Therefore, no excitation has been lost or, equivalently, all the measurements where the number of excitations is $0$ are rejected.
Consequently, we can deduce the time evolution of an observable such as
\begin{equation}\label{Eq:postselection}
    \expect{\hat{\sigma}_+^{(1,\, 2)}\hat{\sigma}_-^{(1,\, 2)}}(t) =\operatorname{Tr}\left [ \hat{\sigma}_+^{(1,\, 2)}\hat{\sigma}_-^{(1,\, 2)} \hat{\rho}_{\rm PS}(t) \right ].
\end{equation}
We remark that Eqs. (\ref{Eq:Post_sel_rho}) and (\ref{Eq:postselection}) is no more linear in the density matrix, since $\rhot$ appears both in the numerator and the denominator.

We plot the results of \eqref{Eq:postselection} in Fig.~\ref{fig:no_local_dissipation}(a) and Fig.~\ref{fig:perfect_transfer}(a) as blue squares for qubit 1 and red circles for qubit $2$. Obviously, \eqref{Eq:postselection} perfectly captures the state transfer, and can be used as a witness of this phenomenon.

We stress that, however, this protocol is not equivalent to that in Sec.~\ref{Sec:Histograms_jumps}. Indeed, in this case one does not collect the quantum jumps of the system, but has to directly effect a PVM to measure the system state.
As such, the method in Sec.~\ref{Sec:Histograms_jumps} uses detectors, while this one needs a simultaneous measure of the total number of excitations in the system.

\section{Fully-dissipative Maxwell's Demon}
\label{Sec: Maxwell_Daemon}

The state transfer protocol can be witnessed also in the presence of thermal processes.
In this case, the system is characterized by six jump operators:
\begin{subequations}
\begin{equation}
 \gamma_1 \left(n_{\rm th}^{(1)} +1 \right) \DD\left[\hat{\sigma}_{-}^{(1)} \right] + \gamma_1 n_{\rm th}^{(1)} \DD\left[\hat{\sigma}_{+}^{(1)} \right]   ,
\end{equation}
\begin{equation}
    \gamma_2 \left(n_{\rm th}^{(2)} +1 \right)  \DD\left[\hat{\sigma}_{-}^{(2)} \right] +  \gamma_2 n_{\rm th}^{(2)} \DD\left[\hat{\sigma}_{+}^{(2)} \right]  ,
\end{equation}
\begin{equation}
     \gamma_c \left(n_{\rm th}^{(c)} +1 \right) \DD\left[\frac{\hat{\sigma}_{-}^{(1)}+ \hat{\sigma}_{-}^{(2)}}{\sqrt{2}}\right] + \gamma_c n_{\rm th}^{(c)} \DD\left[\frac{\hat{\sigma}_{+}^{(1)}+ \hat{\sigma}_{+}^{(2)}}{\sqrt{2}}\right],
\end{equation}
\end{subequations}
where $n_{\rm th}^{(j)}$ represents the temperature of the bath  associated with the $j$-th jump operator.

In the one-excitation manifold, the evolution of the system has exactly the same form of the original problem when introducing the following effective rates
\begin{equation}
\begin{split}
    \gamma_1 & \to \tilde{\gamma}_1= \gamma_1 (1 + 2 n_{\rm th}^{(1)}),
    \\
    \gamma_2 & \to \tilde{\gamma}_2=\gamma_2 (1 + 2 n_{\rm th}^{(2)}),
    \\
    \gamma_c & \to \tilde{\gamma}_c = \gamma_c(1 + 2 n_{\rm th}^{(c)}),
\end{split}    
\end{equation}
as it can be easily argued from \eqref{Eq:Effective_Hamiltonian_counting}.

For the sake of simplicity, let us now assume that the collective bath is at zero temperature [$n_{\rm th}^{(c)}=0$]. 
Since the effective dissipations $\tilde{\gamma}_i$ are determined by the combination of the temperatures (via $n_{\rm th}^{(i)}$) and by the initial dissipation rates $\gamma_i$, we can have $\tilde{\gamma}_1>\tilde{\gamma}_2$ (the condition for the state transfer)
even if $n_{\rm th}^{(2)}>n_{\rm th}^{(1)}$.
Physically speaking, the strongly dissipative bath of qubit $1$ is at a low-temperature (from now on, we will call it the ``cold'' qubit). The qubit $2$ has a bath at a higher temperature that dissipates at a slower rate (the ``hot'' qubit).

We deduce that it is possible to witness a quantum state transfer from the cold qubit to the hot one, which then loses energy to the hot bath.
The process is: (i) The cold qubit absorbs one excitation from the cold bath; (ii) The state transfer induces the hot qubit to become excited as a consequence of jumps not happening; (iii) The hot qubit dissipates in the hot bath.
This is clearly a non-thermal process (which can only happen in an improbable chain of events).
Such a set of events can be witnessed by a counting trajectory protocol if the system is correctly monitored (see Sec.~\ref{Sec:discussion_jumpless_evolution}).

On average, according to the second law of thermodynamics, what will happen is that the hot bath will transfer excitations to the cold bath and to the zero-temperature collective bath.
Clearly, if the collective bath is not connected to the qubits, no transfer can happen, as discussed in Sec. \ref{Sec:discussion_jumpless_evolution}, and there is not heat flow. 

One can now think of connecting the collective bath only when the cold qubit (i.e., qubit 1) is excited.
Within such a configuration, and for sufficiently low coupling $\gamma_2$, the probability that also qubit $2$ is excited is remarkably low.
Thus, three things can happen: (i) The excitation is re-emitted in the cold bath; (ii) The excitation is emitted in the collective bath; (iii) The excitation is emitted in the hot bath.
No processes allow an excitation originating in the hot bath to be emitted in the cold or zero-temperature collective one. In other words, the cold bath is cooling while the hot bath is heating.
This is obviously a single-shot violation of the second-law of thermodynamics and, if repeated enough times, it allows to cool the cold bath by heating the hot one. Furthermore, connecting the collective bath at the right time does not require energy, but only the knowledge that qubit 1 is excited. As such, one is producing a refrigerator by using information \cite{QuanPRL2006,QuanPRE2007,KoskiPRL15,MasuyamaNatComm18,NajeraPRR20}.

This is the definition of a Maxwell's demon \cite{MaruyamaRMP09}, a ``thought'' experiment to demonstrate that the second law is only a statistical principle valid on average and proving the equivalence between entropy and information.
Notice that one continuously needs to measure the cold qubit in order to be assured that the cold system is actually excited, and therefore one is mapping the quantum state of the system into a classical information problem \cite{MaruyamaRMP09}.
Equivalently, a projective measurement on the cold qubit is required, and the collective decay is turned on only if the measurement outcome is ``excited''.
As soon as the cold qubit is excited, the measure is ended and the system evolves under the action of the effective Hamiltonian alone.

Thus, such a Maxwell's demon is purely dissipative in nature, and it is only the quantum backaction deriving from the fact that no quantum jump took place which transfers the excitation.
From an algebraic point of view, the collective dissipation introduces a term of the form $i\gamma_c/2 \left(\ket{e,g}\bra{g,e} +{\rm h.c.} \right)$ in the effective Hamiltonian.
This term has exactly the form of an exchange operator between the two qubits, i.e., the term one would normally induce a coupling.
However, differently from its unitary correspective, this Maxwell's demon does not exchange the excitation with certainty.

In the famous description of the demon opening and closing the door to let slow molecules escape from a hot room (cooling the colder one), the demon actively opens the door ensuring the passage of the particle. 
Here, instead, connecting the collective dissipation is like creating a door, but the finite efficiency of the state transfer means that the door is opened with a finite probability.
This is the reflection of the dissipative and stochastic nature of this Maxwell's demon.
Such an effect can never happen classically, since the state transfer is a purely quantum effect due to quantum superposition induced by a POVM.

\begin{figure}
    \centering
    \includegraphics[width=0.94 \linewidth]{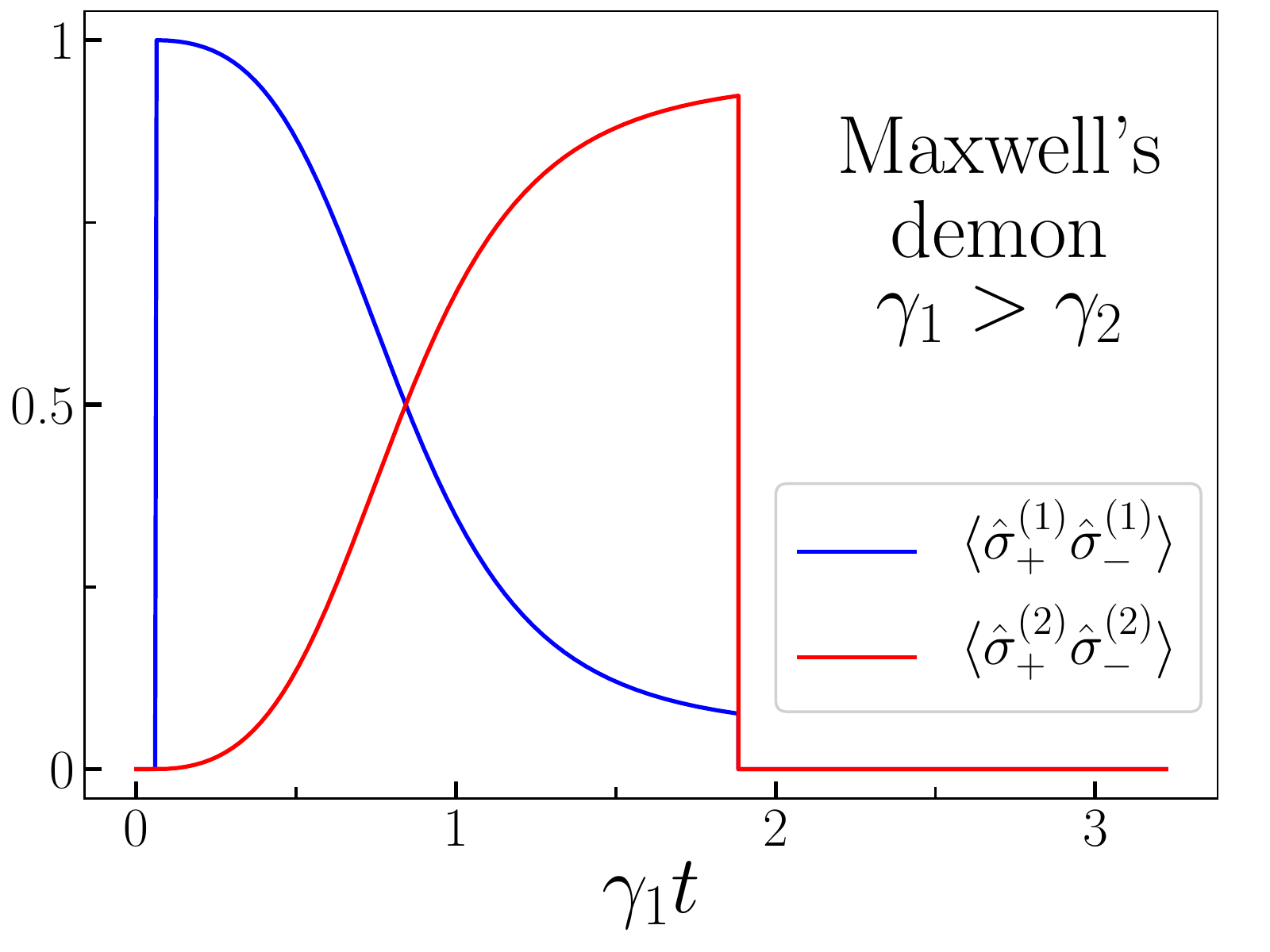}
    \caption{As a function of time, the expecation value of $\hat{\sigma}_+^{(1)}\hat{\sigma}_-^{(1)}$ (for qubit 1, blue curve) and $\hat{\sigma}_+^{(2)}\hat{\sigma}_-^{(2)}$ (for qubit 2, red curve) for an ideal quantum state transfer in the presence of thermal processes, $n_{\rm th}^{(1)}=0.05$, $n_{\rm th}^{(2)}=0.1$, and $n_{\rm th}^{(c)}=0$. Qubit $1$ is at a low-temperature while qubit $2$ has a bath at a higher temperature. The collective dissipation, i.e., the element which allows a state transfer, is only activated ($\gamma_c \neq 0$) once the cold qubit absorbs one excitation, inducing the hot qubit to become excited as a consequence of jumps not happening. Such a Maxwell's demon is purely dissipative in nature. Parameters: $\gamma_2 = \gamma_1/11$, and when active $\gamma_c = 5\gamma_1/11 $ (i.e., when $\gamma_c\neq 0$, the parameters are as in Fig.~\ref{fig:perfect_transfer}). }
    \label{fig:maxwell}
\end{figure}

Figure~\ref{fig:maxwell} shows the time evolution of a system where the Maxwell's demon acted. Initially, the system evolves under the action of the hot and cold baths, but not of the collective dissipation (i.e., $\gamma_c=0$).
The cold qubit (qubit 1) is also continuously monitored to detect if an excitation is present.
As soon as an excitation in the cold qubit is detected ($\gamma_1 t \simeq 0.1$), one stops monitoring the cold qubit and the collective dissipation is activated ($\gamma_c\neq 0$). The quantum state transfer happens and the hot qubit (qubit 2) loses the excitation in the hot bath ($\gamma_1 t \simeq 2$).
After that, $\gamma_c$ is set again to zero, so that if the hot qubit gains an excitation, it can never transfer it to the cold bath.
The overall effect is that the hot bath can only gain energy. 

\section{Conclusions}
\label{Sec:Conclusion}

In this work we described a protocol to transfer excitations from one qubit to another one via purely dissipative non-Markovian processes.
The two qubits are not interacting, but it is the presence of  (both local and collective) dissipation and of the measurement backaction along a single quantum trajectory which allows the transfer to happen.
The process is purely quantum since it relies on an update of the system wave function upon the continuous measurement performed by detectors. Furthermore, the transfer does not involve any coherent process and relies only on the presence of collective and local dissipations.

We derive analytical results using an effective Hamiltonian of the system describing a jumpless evolution upon a counting protocol.
We discuss the possible ways to witness such a dissipative quantum state transfer by making a histogram (considering only local quantum jumps) and via postselection of the jumpless trajectories. In the presence of thermal processes, we demonstrate that one can create a fully dissipative Maxwell's demon by switching on and off the collective dissipation.

All the elements presented in this article can be engineered and realized with state-of-the art quantum optical setups, especially when considering circuit-QED platforms.
The demonstration of the quantum state transfer and of the possibility to realize the fully dissipative Maxwell's demon are a POVM equivalent to the POV standard description of quantum teleportation and Maxwell's demon.
As such, these are effects at the heart of quantum mechanics, based solely on the principles of generalized measures.

The analysis of this system demonstrates that certain behaviours, hidden by the average over many trajectories, acquire a clear and measurable physical meaning at the single trajectory level \cite{BartoloEPJST17,RotaNJP18,MunozPRA19}.
A future perspective is thus to use single trajectories to unveil hidden behaviours in more exotic systems, e.g., those characterized by ultrastrong light matter coupling (USC) regime \cite{kockum2019}.
Beyond the physical interest in investigating these systems, it can be difficult to experimentally observe the presence of phenomena triggered by the high-order light-matter processes typical of USC \cite{GarzianoPRA15,GarzianoPRL16}.
As demonstrated in this article, the statistics of quantum jumps for a properly initialized system can demonstrate the presence of exotic exchanges.

\begin{acknowledgments}
The first two authors (F.M. and V.M) equally contributed to this work.
We thank Ken Funo and Anton F. Kockum for critical reading and suggestions. F.N. is supported in part by: NTT Research, Army Research Office (ARO) (Grant No. W911NF-18-1-0358), Japan Science and Technology Agency (JST) (via the CREST Grant No. JPMJCR1676), Japan Society for the Promotion of Science (JSPS) (via the KAKENHI Grant No. JP20H00134 and the JSPS-RFBR Grant No. JPJSBP120194828), the Asian Office of Aerospace Research and Development (AOARD) (via Grant No. FA2386-20-1-4069), and the Foundational Questions Institute Fund (FQXi) via Grant No. FQXi-IAF19-06.
\end{acknowledgments}

\appendix 
$\ $

\section{Ideal counting trajectory for two qubits interacting with local and collective environment}
\label{Appendix_A}
The evolution along a counting trajectory before a quantum jump takes place is dictated by its effective Hamiltonian.
In the case under consideration, it reads:
\begin{widetext}
\begin{equation}
    \hat{H}_{\rm eff}= \frac{\omega_1 - i  (\gamma_1 + \gamma_c)/2}{2} \hat{\sigma}_z^{(1)} + \frac{\omega_2 - i  (\gamma_2 + \gamma_c)/2}{2} \hat{\sigma}_z^{(2)} -i \frac{\gamma_c}{2} \left(\hat{\sigma}_+^{(1)} \hat{\sigma}_-^{(2)} + \hat{\sigma}_+^{(2)} \hat{\sigma}_-^{(1)} \right).
\end{equation}
Here, $\hat{H}_{\rm eff}$ commutes with the operator $(\hat{\sigma}_z^{(1)}+ \hat{\sigma}_z^{(2)})$. Therefore, the dynamics in the space with one qubit excitation can be decoupled to that of the other excitation manifolds.

Thus, by projecting the time-evolution operator $\hat U(t)={\rm exp} \left( -i \hat{H}_{\rm eff} t \right)$ onto the $2D$-subspace $ \{ |g,e\rangle, |e,g\rangle \} $ we obtain:
\begin{equation}
\begin{split}
\label{evolution_operator} 
\hat U(t)=&\frac{\exp\left(-\Gamma t/4\right) }{ \eta} 
\left \{
\left[ 
\eta \cosh\left( \frac{\eta t}{4}\right) 
- \left( \Delta \gamma+2 i \Delta \omega \right)  \sinh \left( \frac{\eta t}{4}\right)
\right ]  \ket{e,g}\bra{ e,g}
 \right.
- \gamma_c \sinh\left(\frac{\eta t}{4} \right) \left(\ket{g,e}\bra{ e,g}+\ket{e,g}\bra{ g,e} \right)\\
& \qquad \, \, \, \left.
+
\left[ 
\eta \cosh\left( \frac{\eta t}{4}\right) 
+ \left( \Delta \gamma+2 i \Delta \omega \right)  \sinh \left( \frac{\eta t}{4}\right)
\right ]  \ket{g,e} \bra{g,e}
\right\} \, ,
\end{split}
\end{equation} 
where $\Gamma=\gamma_1+\gamma_2+\gamma_C$, $\eta^2=\gamma_C^2 + (\Delta \gamma+2 i \Delta \omega)^2$, $\Delta \gamma=\gamma_1-\gamma_2$, and $\Delta \omega=\omega_1-\omega_2$.

By applying the time-evolution operator $U(t)$ to a system initialized in the state $|\psi (0)\rangle=|e,g\rangle$, we have
\begin{equation}
 \label{initial_state} 
\begin{split}
|\psi (t) \rangle=&\frac{\exp\left(-\Gamma t/4\right) }{ \eta}  \left\{\left[ 
\eta \cosh\left( \frac{\eta t}{4}\right) 
- \left( \Delta \gamma+2 i \Delta \omega \right)  \sinh \left( \frac{\eta t}{4}\right)
\right ]   \ket{e,g} - \gamma_c \sinh\left(\frac{\eta t}{4} \right) \ket{g,e} \right\}.
\end{split}   
\end{equation} 
\end{widetext}

From \eqref{initial_state} we can already see that, in order to have a complete excitation swap at $t \to \infty$, we would need $\eta=\Delta \gamma$. This, however, can be achieved only for $\Delta \omega =0$ (which  can be implemented without any problem) and $\gamma_c\to 0$. The second condition, however, is problematic because it makes the transfer divergently slow, and sooner or later a quantum jump will take place. In this regard, a trade-off between the fidelity of the transfer with respect to $\ket{g,e}$ and the amount of time required to do it is necessary.

\begin{figure}
    \centering
    \includegraphics[width=0.94 \linewidth]{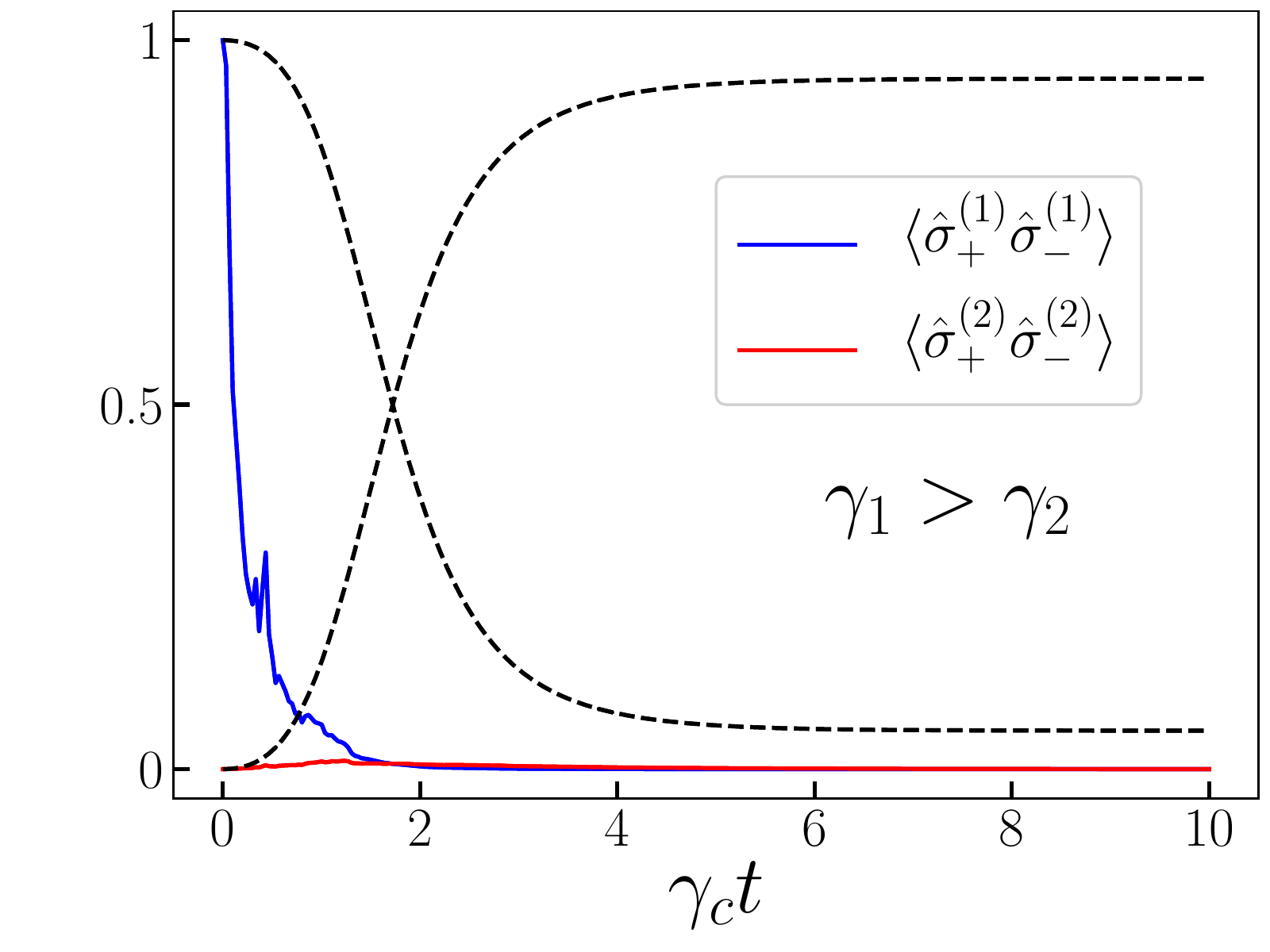}
    \caption{As a function of time, the expecation value of $\hat{\sigma}_+^{(1)}\hat{\sigma}_-^{(1)}$ (for qubit 1, blue curve) and $\hat{\sigma}_+^{(2)}\hat{\sigma}_-^{(2)}$ (for qubit 2, red curve)
    for an ideal homodyne trajectory. This time, the results of a single trajectory differs profoundly from those predicted in \eqref{Eq:Transfer_ideal} (the black dashed lines). The system is initialized in the state $\ket{e, g}$. Parameters as in Fig.~\ref{fig:perfect_transfer}. }
    \label{fig:single_homodyne}
\end{figure}

\begin{figure}
    \centering
    \includegraphics[width=0.94 \linewidth]{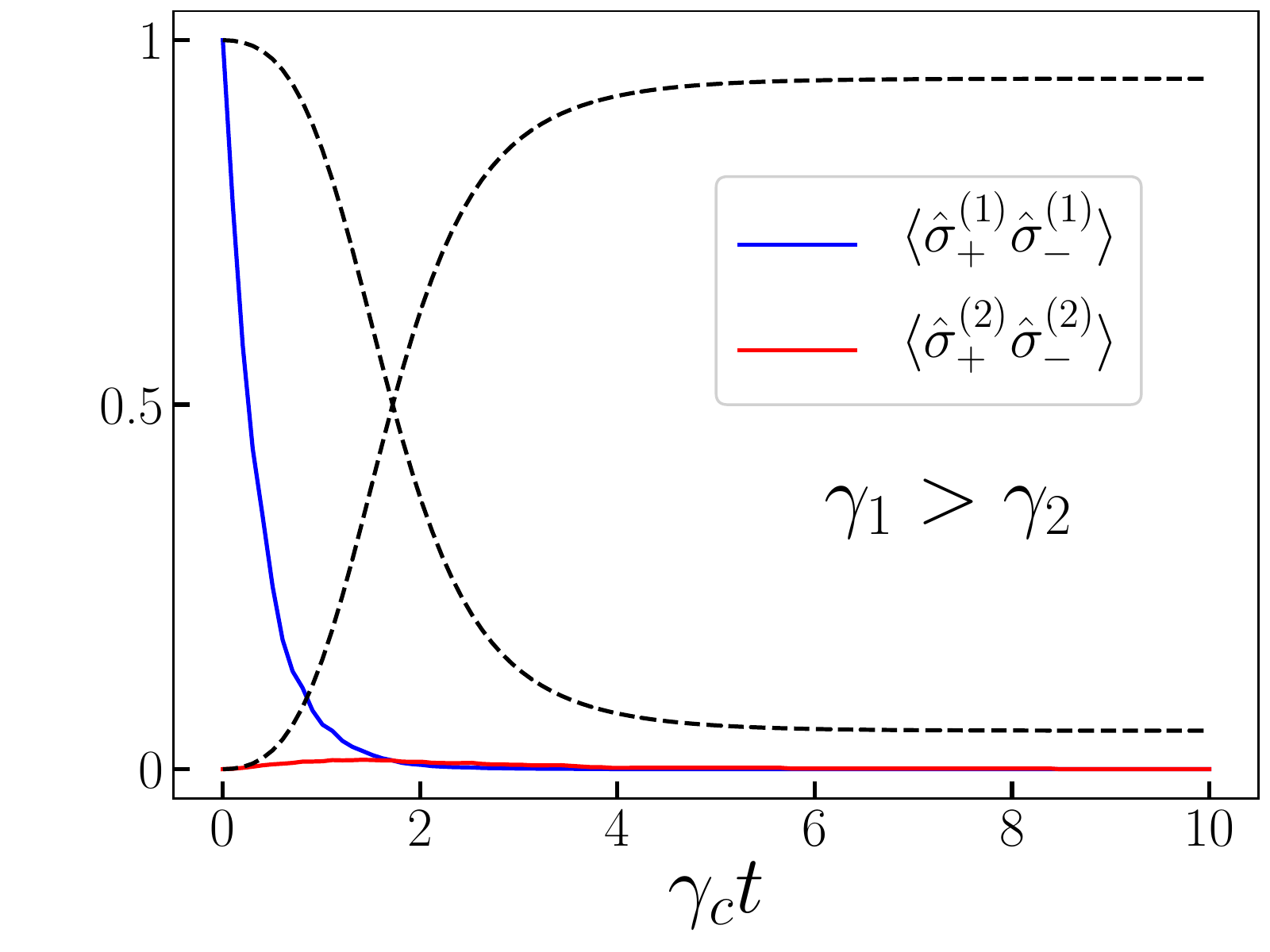}
    \caption{As a function of time, the average over 1000 counting trajectories [i.e., approximately the result of the LME in Eq.~(\ref{Eq:lindblad})] of the expectation value of $\hat{\sigma}_+^{(1)}\hat{\sigma}_-^{(1)}$ (qubit 1, blue curve) and $\hat{\sigma}_+^{(2)}\hat{\sigma}_-^{(2)}$ (qubit 2, red curve). The black dashed lines represent the analytical results of \eqref{Eq:Transfer_ideal}, which is no more predictive. Parameters as in Fig.~\ref{fig:perfect_transfer}. }
    \label{fig:average}
\end{figure}

The dynamics in the 0- and 2-excitation manifolds is trivial, since no evolution (except from the renormalization) occurs according to $\hat{H}_{\rm eff}$. Starting from a state $\ket{e,e}$, no evolution happens until a quantum jump takes place. Whatever jump takes place, the state evolves under the time-evolution operator $\hat U(t)$. If the jump is $\hat{\sigma}_-^{(2)}$, the evolution is the one presented in the main text, since $\ket{\Psi(t_{\rm jump})}=\ket{e,g}$. 

Finally, we stress the importance that the initial state is purely $\ket{e, g}$ to observe the state transfer. Indeed, if the initial wave function is $\ket{\Psi(0)} = A \ket{e, g} + B \ket{g,g}$, the renormalization effect emerging from $\hat U(t)$ will favor the state $\ket{g,g}$. In other words, the state transfer visibility will be attenuated since the ground state becomes more populated as time progresses.

\section{Homodyne trajectories and averaging over many trajectories}
\label{Sec:Homodyne}
To appreciate the importance of the counting protocol for the transfer, let us now consider a single homodyne trajectory.
In this case, a reference laser field is mixed with the output field of each qubit. Consequently, the jump operators become $\hat{J}_\mu (\beta)= \hat{J}_\mu + \beta$, while the Hamiltonian reads
\begin{equation}
    \hat{H}_{\rm eff}(\beta) = \hat{H}_{\rm eff} - i \beta \sum_{\mu} \gamma_\mu \left(\hat{J}_\mu - \hat{J}_\mu^\dagger  \right).
\end{equation}
In the ideal homodyne trajectory limit $\beta\to \infty$, the detector continuously reads a signal. The effect of the latter, however, is minimal, since each quantum jump is largely due to the presence of a local oscillator.
Therefore, the system cannot experience the state transfer, since a measurement always takes place and the back-action does not destroy anymore the state $\ket{\Psi^{+}}$. 

An example of the resulting diffusive trajectory is plotted in Fig.~\ref{fig:single_homodyne}.
We clearly see that the presence of the local oscillator $\beta$ is sufficient to destroy the wanted transfer.
Simulating many homodyne trajectories yields always a similar result, and no state transfer can be witnessed. 

Furthermore, we show that the state transfer is a single trajectory effect which is destroyed by averaging over several trajectories. 
Equivalently, LME approach, which, by its own nature, does not show the state transfer assumes the average over infinitely many quantum trajectories.
This is shown in Fig.~\ref{fig:average}.

\bibliographystyle{FabrizioStyle}
%\bibliography{biblio}


\end{document}